# Single bit full adder design using 8 transistors with novel 3 transistors XNOR gate

Manoj Kumar[1], Sandeep K. Arya[1] and Sujata Pandey[2]

[1]Department of Electronics & Communication Engineering
Guru Jambheshwar University of Science & Technology, Hisar, 125 001, India
[2]Amity University, Noida, 201303, India

## Abstract:

*In present work a new XNOR gate using three transistors has been presented, which shows power dissipation of 550.7272µW in 0.35µm technology with supply voltage of 3.3V. Minimum level for high output of 2.05V and maximum level for low output of 0.084V have been obtained. A single bit full adder using eight transistors has been designed using proposed XNOR cell, which shows power dissipation of 581.542µW. Minimum level for high output of 1.97V and maximum level for low output of 0.24V is obtained for sum output signal. For carry signal maximum level for low output of 0.32V and minimum level for high output of 3.2V have been achieved. Simulations have been performed by using SPICE based on TSMC 0.35µm CMOS technology. Power consumption of proposed XNOR gate and full adder has been compared with earlier reported circuits and proposed circuit's shows better performance in terms of power consumption and transistor count.*

## I. INTRODUCTION

With exponential growth of portable electronic devices like laptops, multimedia and cellular device, research efforts in the field of low power VLSI (very large-scale integration) systems have increased many folds. With the rise in chip density, power consumption of VLSI systems is also increasing and this further, adds to reliability and packaging problems. Packaging and cooling cost of VLSI systems also goes up with high power dissipation. Now a day's low power consumption along with minimum delay and area requirements is one of important design consideration for IC designers. There are three major source of power consumption in CMOS VLSI circuits: 1) switching power due to charging and discharging of capacitances, 2) short circuit power due to current flow from power supply to ground with simultaneous functioning of p-network and n-networks, 3) static power due to leakage currents.

Binary addition is basic and most frequently used arithmetic operation in microprocessors, digital signal processors (DSP) and application-specific integrated circuits (ASIC) etc. Therefore, binary adders are crucial building blocks in VLSI circuits and efficient implementation of these adders affects the performance of entire system. In recent years various types of adder using different logic styles have been proposed [1-12]. Standard CMOS 28 transistor adder using pull up and pull-down network with 14 NMOS transistors and 14 PMOS transistors is most widely reported [1]. In [2] a 16 transistors full adder cell with XOR/XNOR, pass transistor logic (PTL) and transmission gate is reported. Complementary pass-transistor logic (CPL) with 32 transistors having high power dissipation and better driving capability is reported in [4].Transmission gate CMOS adder (TGA) based on transmission gates using 20 transistors is reported in [5]. Main disadvantage of TGA is that it requires double transistors count that of pass transistor logic for implementations of same logic function. A transmission function full adder (TFA) based on transmission function theory used 16 transistors [6]. Multiplexer based adder (MBA) having 12 transistors and elimination of direct path to power supply is reported in [7]. Static energy recovery full (SERF) adder using 10 transistors with reduced power consumption at the cost of higher delay is presented in [8]. Another design of full adder with 10 transistors using XOR/XNOR gates is also reported in [9]. A hybrid CMOS logic style adder with 22 transistors is reported [10]. In [11] a full adder circuit using 22 transistors based on hybrid pass logic (HPSC) is presented. Full adder for embedded applications using three inputs XOR is also reported in [12]. The function of full adder is based on following equation, given three single bit inputs as A, B, $C_{in}$ and it generates two outputs of single bit Sum and $C_{out}$, where:

Sum= $(A \oplus B) \oplus C_{in}$        (1)

$C_{out}$= $A.B + C_{in} (A \oplus B)$        (2)

Structured approach for implementation of single bit full adder using XOR/XNOR has been reported [3] as shown in Figure1. With decomposition of full adder cell into smaller cells, equations (1) and equations (2) can be rewritten as

Sum = H XOR $C_{in}$ = H. $C_{in}$' + H' $C_{in}$     (3)

$C_{out}$ = A. H' + $C_{in}$. H        (4)

Where H is half sum (A XOR B) and H' is complement of H.

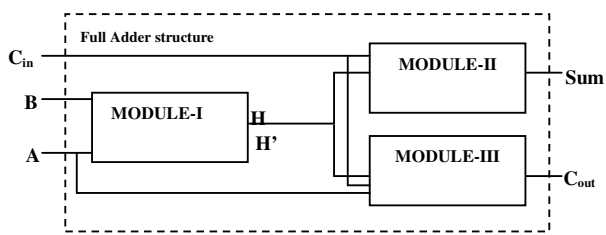

Figure 1: Structure of single bit full adder

The exclusive–OR (XOR) and exclusive–NOR (XNOR) gates are the basic building blocks of a full adder circuit. The XOR/XNOR gates can be implemented using AND, OR, and NOT gates with high redundancy [1]. Optimized design of these gates enhances the performance of VLSI systems as these gates are utilized as sub blocks in larger circuits. XNOR/XOR design with less number of transistors, lesser power dissipation and delay are highly desirable for efficient implementation of the large VLSI system. XOR gate based on eight transistors and six transistors have been used in many earlier designs [1, 13]. Four transistors PTL based XOR gates with limited driving capabilities have been reported in [14]. Different designs for XOR and XNOR gates using four transistors have been presented in [9, 15-17]. Design for three inputs XOR gate instead of two inputs has been reported in [18]. XOR/XNOR design with ten transistors based on transmission gates has been reported in [19]. XOR/XNOR circuits with dual feedback for rectification of degraded logic problem have been reported in [20, 21]. Different power efficient adder designed with body biasing techniques has been reported in [22]. Further, CMOS adder designs with 9 and 10 transistors have been presented in [23].

At circuit level, an optimized design is desirable having less numbers of transistors, small power consumption and adequate output voltage swing. Here, in present work a new XNOR gate with three transistors has been proposed. A single bit full adder having eight transistors based on proposed XNOR module and one multiplexer block having two transistors has been presented. The rest of paper is organized as follows: In Section II, a new three transistors XNOR gate has been reported and single bit full adder circuit based on XNOR gates and multiplexer has been designed. In section III results of proposed XNOR cell and single bit full adder designed in previous section have been presented and compared with previous reported circuits. Section IV concludes the work.

## II. SYSTEM DESCRIPTION

Proposed design of XNOR with three transistors has been shown in Figure 2. In XNOR circuit, the gate lengths of all three transistors have been taken as 0.35μm. Width ($W_n$) of NMOS transistors N1 and N2 has been taken 5.0μm and 1.0μm respectively. Width ($W_p$) for transistor P1 has been taken as 2.0μm. In proposed XNOR, when A=B=0 output is high because transistor P1 is on and N1, N2 transistors are off. With input combinations of A=0 and B=1 circuit shows low output as transistor P1 is off and output node is discharged by transistor N2, which is in on condition. In case when A=B=1, output node shows high logic as transistor N1 is on and high logic is passed to output. Width ($W_n$) of N1 is made large to provide low resistance and due to which signal B is passed to output with less delay, whereas P1 gives higher resistance and higher delay to pass the $V_{dd}$ signal. For N1 transistor width has been increased which provides reduction in threshold voltage as shown in equation (5) [24]. Due to this decrease in $V_t$ the degradation in the output logic is reduced and signal shows sufficient voltage swing.

$$V_t = V_{t0} + \gamma\left(\sqrt{V_{SB} + \phi_0}\right) - \alpha_l \frac{t_{ox}}{L}\left(V_{SB} + \phi_0\right) - \alpha_v \frac{t_{ox}}{L}V_{ds} + \alpha_w \frac{t_{ox}}{W}\left(V_{SB} + \phi_0\right) \qquad (5)$$

Where $V_{t0}$ is the zero bias threshold voltage, $\gamma$ is bulk threshold coefficient, $\phi_0$ is $2\phi_F$ and $\phi_F$ is Fermi potential, $t_{ox}$ is the thickness of oxide and $\alpha_l$, $\alpha_v$ and $\alpha_w$ are the process dependent parameters. From (5) it is obvious that by increasing the W it is possible to reduce the voltage degradation.

Further, due to higher delay provided by PMOS transistor (P1), drain source voltage ($V_{ds}$) of N1 is increased gradually. Transistor N1 operates in non-saturation region and then goes to saturation region. On resistance of MOS transistor consists of the series combination of $R_d$ (drain resistance) and $R_s$ (source resistance) and channel resistance. As for superior switching action higher $V_{gs}$ is desirable. Transistor N1 has larger width, less resistance and higher $V_{gs}$ so switching is very fast in this case. When $V_{ds}$ is less than $V_{gs}$-$V_t$ but greater than zero the channel resistance in non saturation region is given as by [25]

$$R_{on} = \frac{L}{KW(V_{gs} - V_t - V_{ds})} \qquad (6)$$

Here, $R_{on}$ is reduced and signal B is rapidly passed to output which further increase the $V_{ds}$ of transistor N1. In saturation region output resistance $R_{on}$ varies with $V_{ds}$ [26]. As $V_{ds}$ is increased the pinch off moves toward source region and there is incremental change in output impedance [26]. Output resistance is given as

$$R_{on} = \frac{2L}{1 - \frac{\Delta L}{L}} \frac{1}{I_d} \sqrt{\frac{qN_b}{2 \in_{si}} (V_{ds} - V_{ds,sat})} \qquad (7)$$

Where L is the channel length, $\Delta L$ is the small change in length due to increased $V_{ds}$, $I_d$ is the drain current, and $N_b$ is doping concentration. With higher $V_{ds}$, the output resistance increase in saturation region. This condition is applicable for both N1 and N2 because of high $V_{ds}$ for these transistors. Due to higher resistance provided by N1 and N2 in saturation region the high output logic is maintained at output node.

In another case when A=1 and B=0 both transistors (P1 & N1) are on and output node is discharged rapidly by N1 and N2 transistors. In this case with A=1 transistor N1 turns on which further turn on the transistor N2 and a conducting path is provided by N1 and N2. This connectivity of output node with ground discharges the output node. The switching speed of N1 is higher than N2 because delay is inversely proportional to channel width [1]. Due to on condition of transistor N1 the gate voltage of N2 increase above its threshold voltage and transistor N2 also goes in on condition. In this position the circuit is just behaving like an inverter with A=1 as input and gives output as low logic. Transistor P1 is just acting as load resistance with grounded gate input (B= 0).

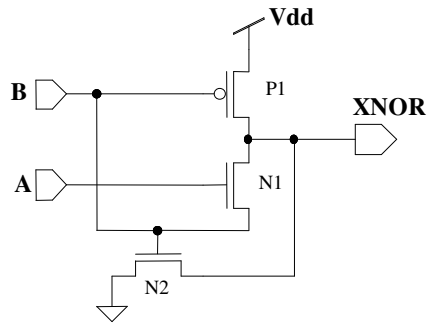

Figure 2: Design of proposed 3 transistors XNOR gate

XOR operation can be obtained with addition of additional inverter. Signals levels are also restored to $V_{dd}$ in the proposed design by addition of inverter with little increase in power consumption. Complete XNOR/XOR module with five transistors has been shown in Figure 2. Typical values of transistor widths ($W_p$=2.0μm) for P2 and ($W_n$=1.0μm) for N3 have been taken for inverter section. Simulations of XNOR gate and XNOR/XOR cell also have been carried out with varying supply voltage from [3.3 - 1.8] V.

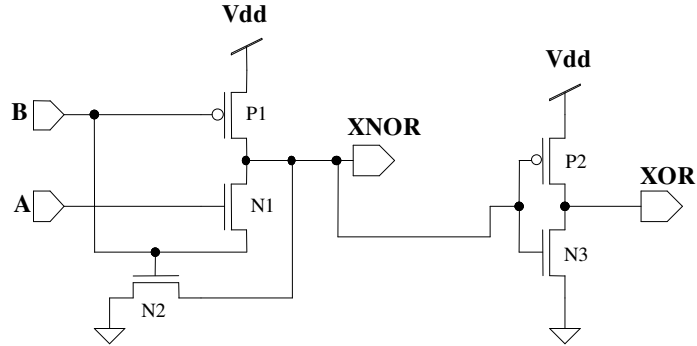

Figure 3: Design of proposed XNOR/XOR cell

Full adder circuit can be implemented with different combinations of XOR/XNOR modules and two multiplexer [2, 17] but this approach has not been used in current work as proposed XNOR/XOR cell shows high power consumption than single XNOR gate. Proposed full adder circuit has been implemented by two XNOR gates and one multiplexer block as shown in block diagram of Figure 4(a). Sum is generated by two XNOR gates and $C_{out}$ is generated by two transistors multiplexer block. The single bit full adder using proposed XNOR gates with eight transistors has been implemented and shown in Figure 4(b). For multiplexer section typical values of width ($W_n$ & $W_P$) 1.0µm & 2µm for NMOS and PMOS transistors have been taken with gate length of 0.35µm. Simulations have been performed using SPICE based on TSMC 0.35µm CMOS technology with supply voltage of 3.3V.

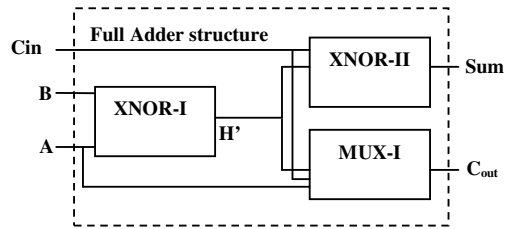

(a)

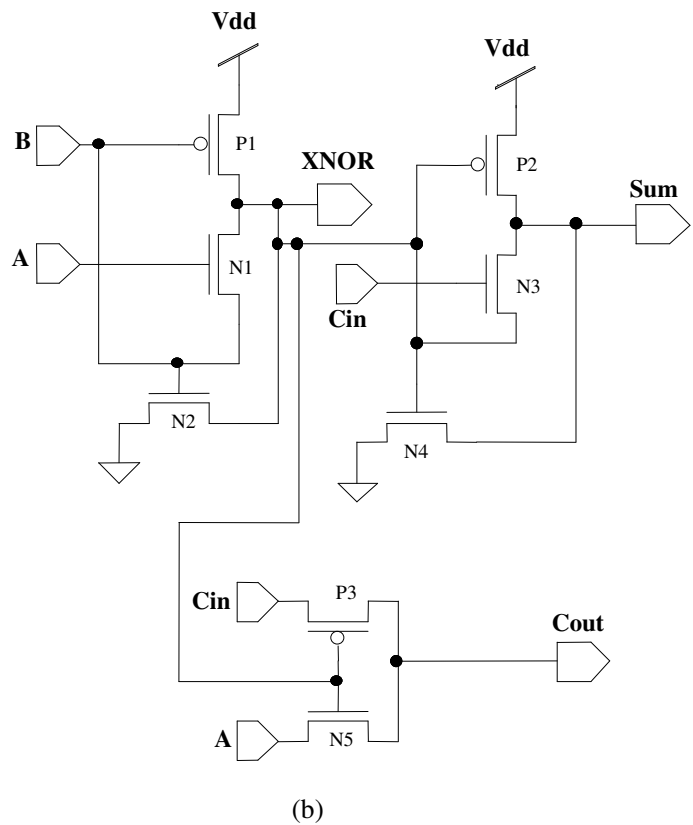

(b)

Figure 4: Full adder using two XNOR gates and multiplexer (a) Block diagram
(b) Circuit diagram

## III. RESULTS AND DISCUSSIONS

Table 1 shows power consumption, delay and output voltage levels with varying supply voltage [3.3 -1.8] V for proposed XNOR gate [Figure 2]. Power consumption varies from [500.727 - 89.931] μW with variations in supply voltage from [3.3 - 1.8] V as shown in Figure 5. Delay varies from [14.466 - 23.050] ps with variations in supply voltage from [3.3 - 1.8] V and has been shown in Figure 6. Power consumption has been reduced due to reduced transistor count and reduced capacitance in the XNOR circuit. Short circuit currents are also quite low in the circuit as direct path from supply to ground is eliminated. It has been observed from Figure 5 & 6 that power consumption reduces, whereas delay increase with reduction in supply voltage. Table I also show that proposed XNOR gate provide sufficient output voltage levels and noise margin of approximately 2V has been obtained with 3.3V supply.

Table 1: Power consumption, delay and output voltage levels of proposed XNOR gate

| Supply voltage (V) | Power consumption (µW) | Output delay (ps) | Minimum level for high output (V) | Maximum level for low output (V) |
|---|---|---|---|---|
| 3.3 | 500.727 | 14.466 | 2.05 | 0.084 |
| 3.0 | 395.378 | 15.846 | 1.84 | 0.072 |
| 2.7 | 301.417 | 17.269 | 1.63 | 0.0648 |
| 2.4 | 218.996 | 18.165 | 1.41 | 0.05418 |
| 2.1 | 148.329 | 20.083 | 1.20 | 0.04339 |
| 2.0 | 127.452 | 21.099 | 1.13 | 0.04145 |
| 1.8 | 89.931 | 23.050 | 0.92 | 0.03400 |

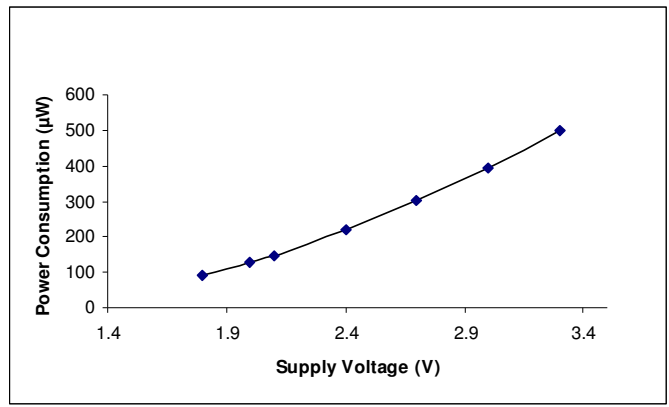

Figure 5: Power consumption of XNOR gate with supply voltage

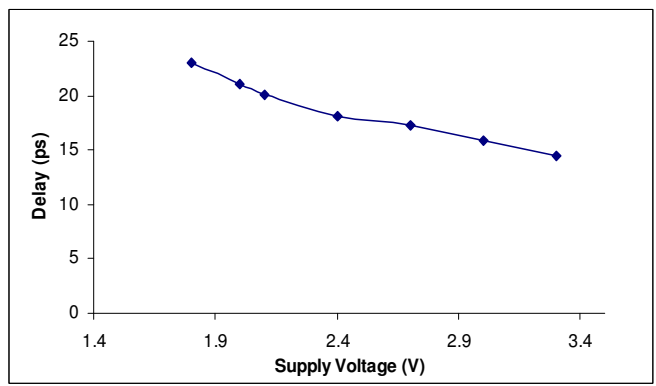

Figure 6: Delay of XNOR gate with supply voltage

Table 2 show results of power consumption and output voltage levels for XNOR/XOR cell

[Figure 3] with variations in supply voltage from [3.3 - 1.8] V. Power consumption is increased with addition on CMOS inverter and XNOR/XOR cell shows much higher power consumption than single XNOR gate. Minimum level for high output for XOR has also been increased with addition of inverter. Noise margin of circuit is approximately 2V which is near to previous case of XNOR gate. Figure 7 show variations of power consumption for XNOR/XOR cell with supply voltage. Figure 8 shows input and output waveform results for XNOR/XOR cell.

Table 2: Power consumption and output voltage levels of proposed XNOR/XOR cell

| Supply voltage (V) | Power consumption (μW) | Minimum level for high output (V) | Maximum level for low output (V) |
|---|---|---|---|
| 3.3 | 1104.8 | 2.5 | 0.69 |
| 3.0 | 828.38 | 2.3 | 0.65 |
| 2.7 | 666.59 | 2.17 | 0.61 |
| 2.4 | 396.94 | 1.82 | 0.55 |
| 2.1 | 241.79 | 1.56 | 0.51 |
| 2.0 | 199.02 | 1.47 | 0.50 |
| 1.8 | 127.02 | 1.30 | 0.455 |

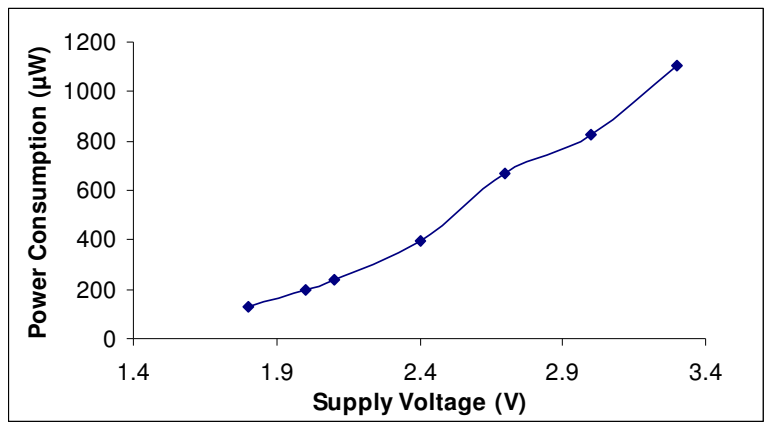

Figure 7: Power consumption of XNOR/XOR cell with supply voltage

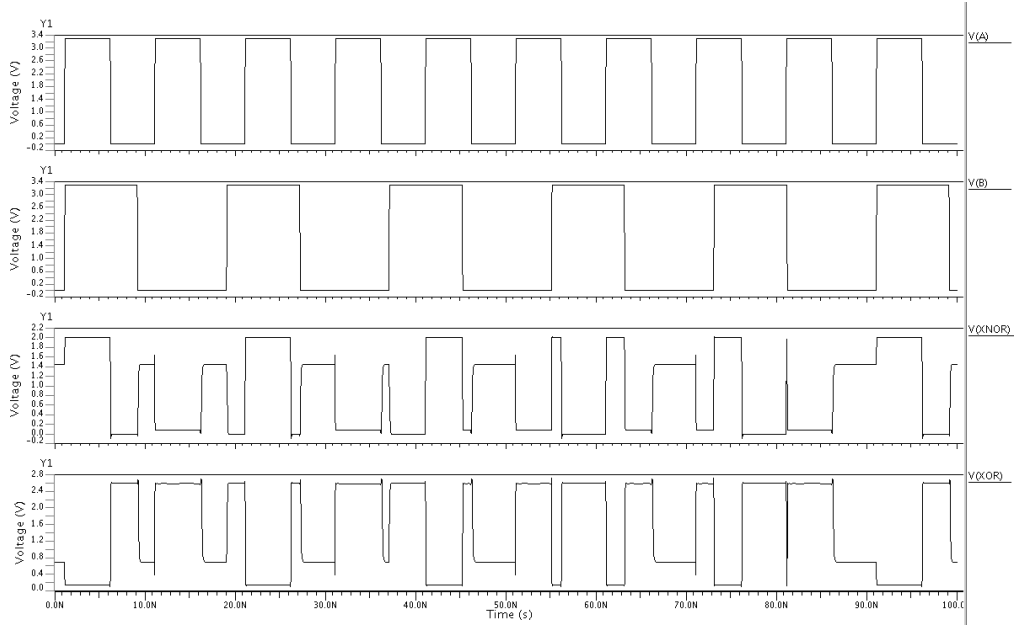

Figure 8: Input and out waveform for XNOR/XOR cell

Table 3 shows power consumption, output delay and output voltage levels for sum and carry signals for eight transistors single bit adder. Proposed adder circuit shows power consumption of 581.542 µW with acceptable level of output. Figure 9 shows the input and output waveform results for full adder circuit. Proposed full adder has less internal capacitance due to reduce number of transistors and shows reduced power consumption. A wide range of simulation has been done form 3.3 V to 1.8 V to see levels of output signal circuit which shows acceptable voltage levels are obtained with proposed circuits.

Table 3: Performance of proposed full adder

| Power Consumption (µW) | 581.542 |
|---|---|
| Sum delay (ps) | 15.1311 |
| Carry delay (ps) | 3.372 |
| Minimum level for high sum output (V) | 1.97 |
| Maximum level for low  sum output (V) | 0.24 |
| Minimum level for high carry output (V) | 3.2 |
| Maximum level for low carry  output (V) | 0.32 |

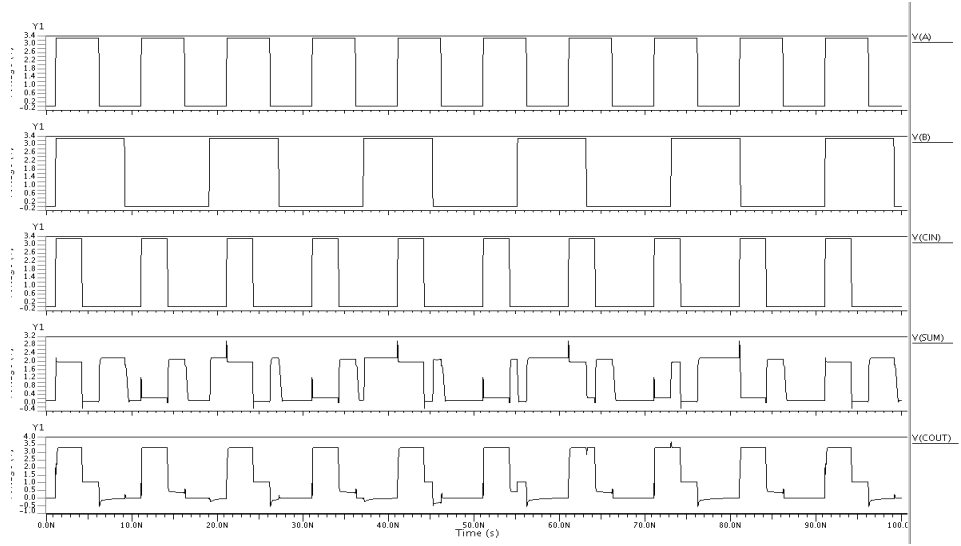

Figure 9: Input and output patterns for proposed full adder

Table 4 shows comparisons of power consumption, output voltage levels for XNOR gate in present work and earlier reported circuits. Graph drawn in Figure 10 shows that proposed XNOR cell has lowest power consumption among compared circuits.

Table 4: Comparisons of power consumption of proposed XNOR Cell

| XNOR Configuration | Power consumption(µW) | Minimum level for high output (V) | Maximum level for low output (V) |
|---|---|---|---|
| 4T[14] | 587.63 | 1.86 | 0.0006 |
| 4T[17] | 818.39 | 1.76 | 0.0005 |
| 4T[15] | 1059.2 | 1.91 | 0.39 |
| 8T[1] | 917.66 | 1.92 | 0.0004 |
| 3T (present work) | 500.72 | 2.05 | 0.084 |

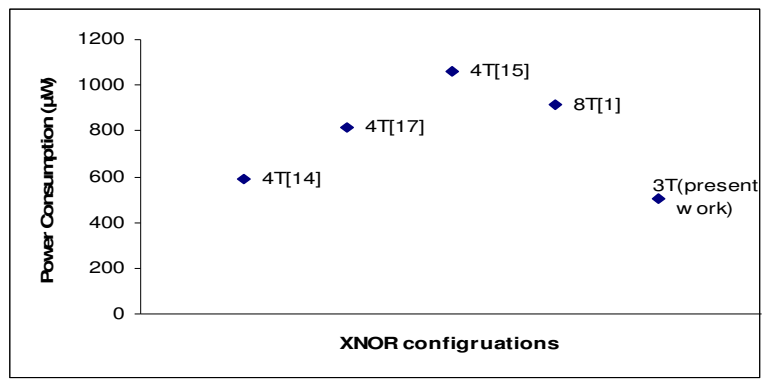

Figure 10: Power consumptions comparisons for XNOR gates

Adders reported in [2], [5], [6], [8], [10], [11] have been simulated and comparisons have been presented in Table 5. Graphs drawn in Figure 11 shows comparisons of power consumption of proposed circuit and earlier reported circuits. It has been shown that proposed adder show less power consumption than previously reported adders except ten transistors SERF adder [8]. Proposed circuits also show superiority in terms of transistor count with earlier reported circuits.

Table 5: Power consumption comparisons of proposed full adder

| Adder configuration | Power consumption(μW) | Number of transistors for design |
|---|---|---|
| TGA20T [6] | 1255.54 | 20 |
| 16T adder [2] | 591.07 | 16 |
| 10T SERF [8] | 531.29 | 10 |
| 22T hybrid adder [10] | 1836.4 | 22 |
| 22T HPSC [11] | 1533.9 | 22 |
| 18T [5] | 617.23 | 18 |
| 8T [present work] | 581.542 | 08 |

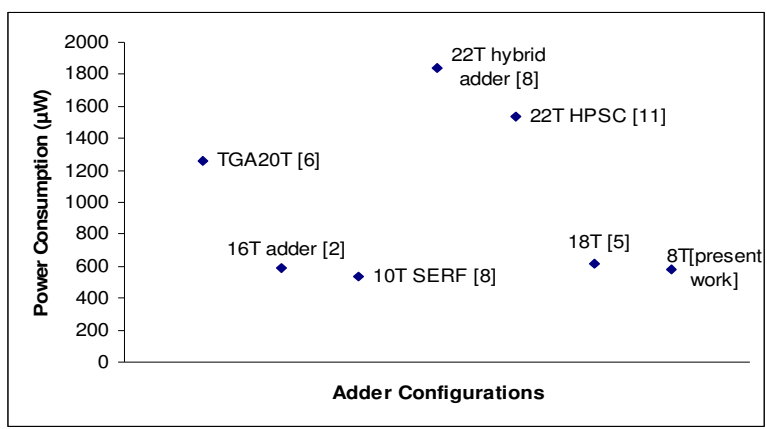

Figure 11: Power consumptions comparisons of proposed adder with earlier reported circuits

## IV. CONCLUSIONS

In current work, a new low power XNOR gate with three transistors have been reported which shows power dissipation of 500.7272 μW. Compared with earlier reported XNOR gates, proposed circuit shows less power consumption and better output signal levels with reduce transistor count. A single bit full adder with 8 transistors based on proposed XNOR gate has been presented which show power consumption of 581.542μW with maximum output delay of 15.1311 ps. Proposed full adder has been compared with earlier reported circuits and reported circuit shows reduced power consumption with less number of transistors.